\documentclass[useAMS,fleqn, usenatbib]{mn2e}
\usepackage{amsmath}%
\usepackage{amsfonts}%
\usepackage{amssymb}%
\usepackage{graphicx}
\usepackage{times}
\usepackage{hyperref}
\usepackage{color}

\begin{document}
\title[Lorentz-boost effects on shear and convergence]{Effect of Our Galaxy's Motion on Weak Lensing Measurements of Shear and Convergence}

\author[Mertens J.~B., Yoho A., and Starkman G. D.]{J.~B.~Mertens,$^1$ A.~Yoho,$^{1,2}$ 
G.~D.~Starkman,$^{1,2}$ \\
$^1$CERCA/ISO, Department of Physics, Case Western Reserve University,
10900 Euclid Avenue, Cleveland, OH 44106-7079, USA\\
$^{2}$ CERN, CH-1211 Geneva 23, Switzerland}

\maketitle
%%%%%%%%%%%%%%%%%%%%%%%%%%%%%%%%%%%%%
\begin{abstract}
%%%%%%%%%%%%%%%%%%%%%%%%%%%%%%%%%%%%%
In this work we investigate the effect on weak-lensing shear and convergence
measurements due to distortions from the Lorentz boost induced by our Galaxy's motion.
While no ellipticity is induced in an image from the Lorentz boost to first order in 
$\beta\equiv v_{\text{Galaxy}}/c$, the image is magnified.
This affects the inferred convergence at a 10 per cent level, and is most notable for low
multipoles in the convergence power spectrum $C^{\kappa\kappa}_{\ell}$ and  for surveys with
large sky coverage like LSST and DES. Experiments which image only small fractions of
the sky and convergence power spectrum determinations at $\ell\gtrsim 5$ can safely
neglect the boost effect to first order in $\beta$.
\end{abstract}


%%%%%%%%%%%%%%%%%%%%%%%%%%%%%%%%%%%%%
\section{Introduction}
%%%%%%%%%%%%%%%%%%%%%%%%%%%%%%%%%%%%%

Current and upcoming weak-lensing experiments  will map the gravitational potential of structures
out to moderate redshift over most of the sky, thereby enabling us to learn more about the 
nature and distribution of dark matter and dark energy.
To do so, these surveys will measure one to two-per cent changes in the intrinsic shape of surrounding background galaxy images. 
These small image distortions are probes of the 
gravitational potentials of lensing  clusters and thus of the  large-scale structure of our universe. 

Weak lensing therefore provides an indirect method to measure the  distribution of the dark matter and possibly
infer its properties, even though it has yet to be seen in direct detection experiments~\citep{Wright:1999jc}. 
The details of cluster formation also depend on the  density and properties of dark energy,
so  measuring the number of weak lensing clusters as a function of redshift 
will lead to an improved determination of the dark energy equation of state~\citep{Schneider:2005ka}. 
Similarly, measurements of cluster gravitational potentials over a range of redshifts 
are an important component of tests for modifications to General Relativity that might explain cosmic
acceleration and replace dark energy \citep{Lue:2004rj, Bean:2010zq}.

Since weak-lensing surveys are attempting to measure only one per cent distortions 
of the unknown intrinsic shapes of distant objects, observations and data analysis  
can easily be affected by systematic effects. Therefore,
it is important to fully understand and characterize any unwanted signal that
may hinder drawing accurate conclusions from the data. Several systematics have been investigated
to-date, such as~\citep{Cunha:2012us,Chang:2012cp,Yoo:2012vm,Chang:2012cn} among others, however one in particular,
the Lorentz boost of photons caused by our Galaxy's peculiar motion, has been neglected.
In this work we quantify its importance to weak-lensing surveys. This work is complementary to \citep{Bonvin:2008ni} in which the effects of source and observer motion due to their local gravitational fields were considered.

The paper is organized as follows: section~\ref{wl_background} briefly reviews
weak lensing shear and convergence; section~\ref{peculiar}   discusses how Lorentz boosts distort images;
section~\ref{shear} presents
the boosted shear matrix $A_{ij}$; in section~\ref{powerspectrum} we calculate the convergence
power spectrum $C_{\ell}^{\kappa\kappa}$; in section~\ref{reduced} we show the effect of 
boosts on reduced shear. We discuss our results in section~\ref{discussion}.

%%%%%%%%%%%%%%%%%%%%%%%%%%%%%%%%%%%%%%%
\section{Weak Lensing Shear and Convergence}
\label{wl_background}
%%%%%%%%%%%%%%%%%%%%%%%%%%%%%%%%%%%%%%%

As photons from distant galaxies travel towards us, they  traverse the gravitational
potentials of nearby galaxy clusters, causing their paths  to bend. This well-known phenomenon of
gravitational lensing causes the shapes of  observed galaxies to be distorted and the locations of their
images to be offset.  The location  $\theta_s^i$ of a source
 is related to that of its  lensed image, $\theta^i$ 
 (here $i=1,2$ denotes the  component of the image or source location in the plane
 perpendicular to the line-of-sight, a.k.a. the image-source plane):
\begin{equation}\label{theta}
\theta_s^i=\theta^i+\frac{2}{\chi}\int_0^\chi d\chi^\prime \frac{d\Phi(\chi^{\prime})}{dx}\left(\chi - 
\chi^\prime \right).
\end{equation}

The angular shift and image distortion are typically  encoded in a $2\times 2$ 
symmetric transformation matrix
\begin{equation}\label{matrix}
A_{ij}\equiv \frac{\partial \theta_s^i}{\partial \theta^j}\equiv 
\begin{pmatrix}
1-\kappa-\gamma_1 & -\gamma_2 \\
-\gamma_2 & 1-\kappa+\gamma_1
\end{pmatrix}.
\end{equation}
Here $\kappa$ is the convergence, which describes the magnification of a galaxy image, 
and $\gamma_{1,2}$ are components of the shear matrix, which characterizes the stretching and angular 
deflection of the image. 
 
Since surveys are unable to measure shear directly, they instead measure ellipticity of galaxy 
images. Starting with   the quadrupole moment of an image,
\begin{equation}
q_{ij}\equiv \int d^2\theta I_{obs}(\theta)\,\theta_i \theta_j ,
\end{equation}
two standard measures of image ellipticity  are 
\begin{equation}
\epsilon_1 = \frac{q_{xx}-q_{yy}}{q_{xx}+q_{yy}}
%\end{equation}
%and
%\begin{equation}
\quad {\rm and} \quad
\epsilon_2 = \frac{2q_{xy}}{q_{xx}+q_{yy}}\,.
\end{equation}
Clearly $\epsilon_1 = \epsilon_2 = 0$ for a circular image. In the weak 
lensing limit ($\kappa, \gamma\ll1$),  ellipticity and shear are simply related, for example  $\epsilon_1=2\gamma_1$.
We can use this relation to infer the underlying gravitational potential of a cluster
by measuring the shapes (ellipticities) of galaxy images that have been stretched through  weak 
 lensing by the potential of that cluster.

Intrinsic ellipticities of galaxies are 
approximately randomly oriented
 (especially if the galaxies are widely separated in redshift), however
the ellipticites induced by gravitational lensing are correlated when they are nearby on the sky.
Two  galaxy images with their ellipticities oriented in the same direction 
will have the same sign for  $\epsilon_1$, whereas images with anti-aligned ellipticities will have 
opposite signs for $\epsilon_1$. Therefore, two galaxies at points $\theta_1$ and $\theta_2$ with random alignments
(e.g. images not affected by gravitational lensing) will  have 
$\epsilon_1(\theta_1)\epsilon_1(\theta_2) $ negative as often as  positive.  Lensed galaxy pairs 
will be biased toward positive values, when $\theta_1$ and $\theta_2$ are sufficiently close.
This  correlation function can therefore be used statistically as a tool to infer the underlying lensing
potential, assuming sufficient tracer galaxies are lensed. 
 
To determine the convergence $\kappa$, one measures the local number density of galaxies. 
Since magnification changes the size of a galaxy image as well as its brightness,
counts will be lower than expected if the magnification $\mu > 1$. We can again use
the weak lensing limit to find a relation between magnification and convergence. Letting
$\gamma \equiv \sqrt{\gamma_1^2+\gamma_2^2}$:
\begin{equation}
\mu = \frac{1}{det A} = \frac{1}{(1-\kappa)^2-\gamma^2}\simeq1+2\kappa .
\end{equation}
It is therefore important for calculations
of the weak lensing convergence power spectrum to correctly determine the magnification.
Similar to ellipticity, in  actual observations one can measure the reduced shear, $g$,
rather than the shear itself, $\gamma$.  
On the one hand
\begin{equation}\label{rs}
g\equiv\frac{a/b-1}{a/b+1}\,,
\end{equation}
where $a/b$ is the ratio of the semimajor ($a$) and semiminor ($b$)
axes of a galaxy image~\citep{Schneider:2005ka}.
On the other hand $g$  can be expressed in terms
of $\kappa$ and $\gamma$:
\begin{equation}\label{rs}
g=\frac{\gamma}{1-\kappa}.
\end{equation}

It should be clear that any systematic which would be coherent across a patch of sky necessarily
needs to be removed. We will show in the following sections that the Lorentz boosting of galaxy images will
do just that: it affects galaxies at iso-latitute rings around the boost direction
in the same way, which could give a false measurement of correlated ellipticities and convergence 
and, therefore, lead to an incorrect reconstruction of the gravitational lensing potential.

%%%%%%%%%%%%%%%%%%%%%%%%%%%%%%%%%%%%%%%%%
\section{Effect of Peculiar Motion on Magnification and Ellipticity}
\label{peculiar}
%%%%%%%%%%%%%%%%%%%%%%%%%%%%%%%%%%%%%%%%%%%

To obtain a first estimate of how images are distorted due to Lorentz boosts, we 
consider two spatial vectors, $\hat{n}_1$ and $\hat{n}_2$, separated by a small 
angle $\cos\delta\alpha = \hat{n}_1\cdot\hat{n}_2$. 
%We will adopt coordinates in which our motion is in the  ${\bf\hat{z}}$ direction and use spherical 
%coordinates $(\theta,\phi)$, with $\theta$ the polar
%angle measured from ${\bf\hat{z}}$ and $\phi$ the azimuthal angle. 
%
Under a boost $\beta \hat{v}$:
\begin{eqnarray}
    \hat{n}' & = &
    \left( \frac{\cos\theta+\beta}{1+\beta \cos\theta} \right) \hat{v}
    + \frac{\hat{n}-\hat{v}\cos\theta}{\gamma(1+\beta\cos\theta)} \\
    & \simeq & \hat{n} + \beta \left( \hat{v} - \hat{n}\cos\theta \right),
\end{eqnarray}
where $\cos\theta = \hat{n}\cdot\hat{v}$.  For the weak lensing systems under consideration
the size of the lensed field is small, so we are interested in  $\hat{n}_1 \simeq \hat{n}_2$.
Both vectors will then be, to lowest order in small quantities, 
the same angle $\theta$ away from the boost direction. 
This gives:
\begin{eqnarray}
    \cos\delta\alpha'& \!\!\!\!\!=\!\!\!\!\! & \hat{n}_1'\cdot\hat{n}_2' \\
    &\!\!\!\!\! \simeq \!\!\!\!\!& \left( \hat{n}_1 + \beta \left( \hat{v} - \hat{n}_1\cos\theta \right) \right)
    \left( \hat{n}_2 + \beta \left( \hat{v} - \hat{n}_2\cos\theta \right) \right) \\
    &\!\!\!\!\! =\!\!\!\!\! & \cos(\delta\alpha) + 2\beta\cos\theta\left( 1-\cos\delta\alpha \right),
\end{eqnarray}
or, expanding again in powers of $\delta\alpha$,
\begin{equation}
    \delta\alpha' \simeq \delta\alpha \left( 1 - \beta\cos\theta \right).
\end{equation}
Because this equation holds irrespective of the direction $\hat{n}_1-\hat{n}_2$, it follows that,
to first order in the small quantities $\beta$ and $\delta\alpha$,
objects will only undergo magnification.

Any object with a circular cross section
will retain that circular cross section to all orders in $\beta$.
Therefore, for intrinsically circular galaxy images  {\it{no ellipticity will 
be induced due to our motion}}.
(However, circular isophotes may map to elliptical isophotes -- an effect
we have not yet fully investigated, but on which we elaborate in sec.~\ref{discussion}.)
More generally,  any intrinsically elliptical image will only be magnified, 
not deformed, to first order in $\beta$. Higher order
$\beta$ corrections not considered here may produce shape 
distortions. 

Since the images will be magnified at first order, measurements
of convergence {\it will} be affected by relativistic aberration.

%%%%%%%%%%%%%%%%%%%%%%%%%%%%%%%%%%%%%%%%%%%%%%%%%%%%%%%%%%%%%%%%
\section{Boosting the Shear Matrix}
\label{shear}
%%%%%%%%%%%%%%%%%%%%%%%%%%%%%%%%%%%%%%%%%%%%%%%%%%%%%%%%%%%%%%%%

We now present a general procedure for boosting the shear matrix in order to extend the results of the previous section to higher order.

Consider a unit vector  in an arbitrary direction $(\theta,\phi)$ specified in spherical coordinates.  Without loss of genearlity, we choose our coordinate axes such that 
$(\theta,\phi)\simeq(\pi/2-\delta\theta,\pi/2-\delta\phi)$.   In Cartesian coordinates,
\begin{equation}
{\bf x}=\left(\begin{array}{c}
x\\
y\\
z
\end{array}\right)\simeq\left(\begin{array}{c}
-\delta\phi\\
1-\frac{\delta\theta^{2}}{2}-\frac{\delta\phi^{2}}{2}\\
-\delta\theta
\end{array}\right).
\end{equation}
In order to tranform to a frame where the boost is in the $\hat z$ direction, we rotate this unit vector by an arbitrary angle $\psi$ about the ${\bf x}$-axis. Since the boost will not affect the azimuthal angle, there is no need to also rotate about the ${\bf z}$-axis.

We next boost the unit vector in the z-direction, and then rotate back to the original frame.
The boosted coordinates $(\theta',\phi')$, are
again represented as
 $(\pi/2-\delta\theta',\pi/2-\delta\phi')$:
\begin{eqnarray}
\delta\theta' & \simeq &\delta\theta+\beta\delta\theta\sin\psi\,,\\
\delta\phi' & \simeq & \delta\phi+\beta\delta\phi\sin\psi.
\end{eqnarray}
Here we have evaluated everything to second order in small parameters
$\delta\theta$, $\delta\phi$, and $\beta$, and have ignored higher order contributions.

We can immediately see  that there will be induced magnification due
to relativistic aberration. From eq.~\ref{matrix} we have
\begin{equation}
A_{ij} = \left(\begin{array}{cc}
1+\beta\sin\psi & 0\\
0 & 1+\beta\sin\psi
\end{array}\right),
\end{equation}
from which we conclude that only magnification effects are present
at first order in $\beta$: $\kappa = -\beta\sin\psi$. As current measurements of
$\kappa$ from weak lensing surveys are of order $10^{-2}$ and $\beta\sim10^{-3}$,
we expect this magnification to be up to $10$ per cent of weak lensing effects.

%%%%%%%%%%%%%%%%%%%%%%%%%%%%%%%%%%%%%%%%%%%%%%%%%%%%
\section{Calculating the Boosted Convergence Power Spectrum}
\label{powerspectrum}
%%%%%%%%%%%%%%%%%%%%%%%%%%%%%%%%%%%%%%%%%%%%%%%%%%%%
Here we look at the effect of boosting on the convergence power spectrum.
We note that as boosting will (to first order) produce a dipole on
the sky in the observed magnification of sources, we choose to perform
a full-sky decomposition in terms of spherical harmonics, similar
to the CMB. We then look at the power spectrum on cut skies of various
sizes.

\begin{figure}
	\includegraphics[scale=.23]{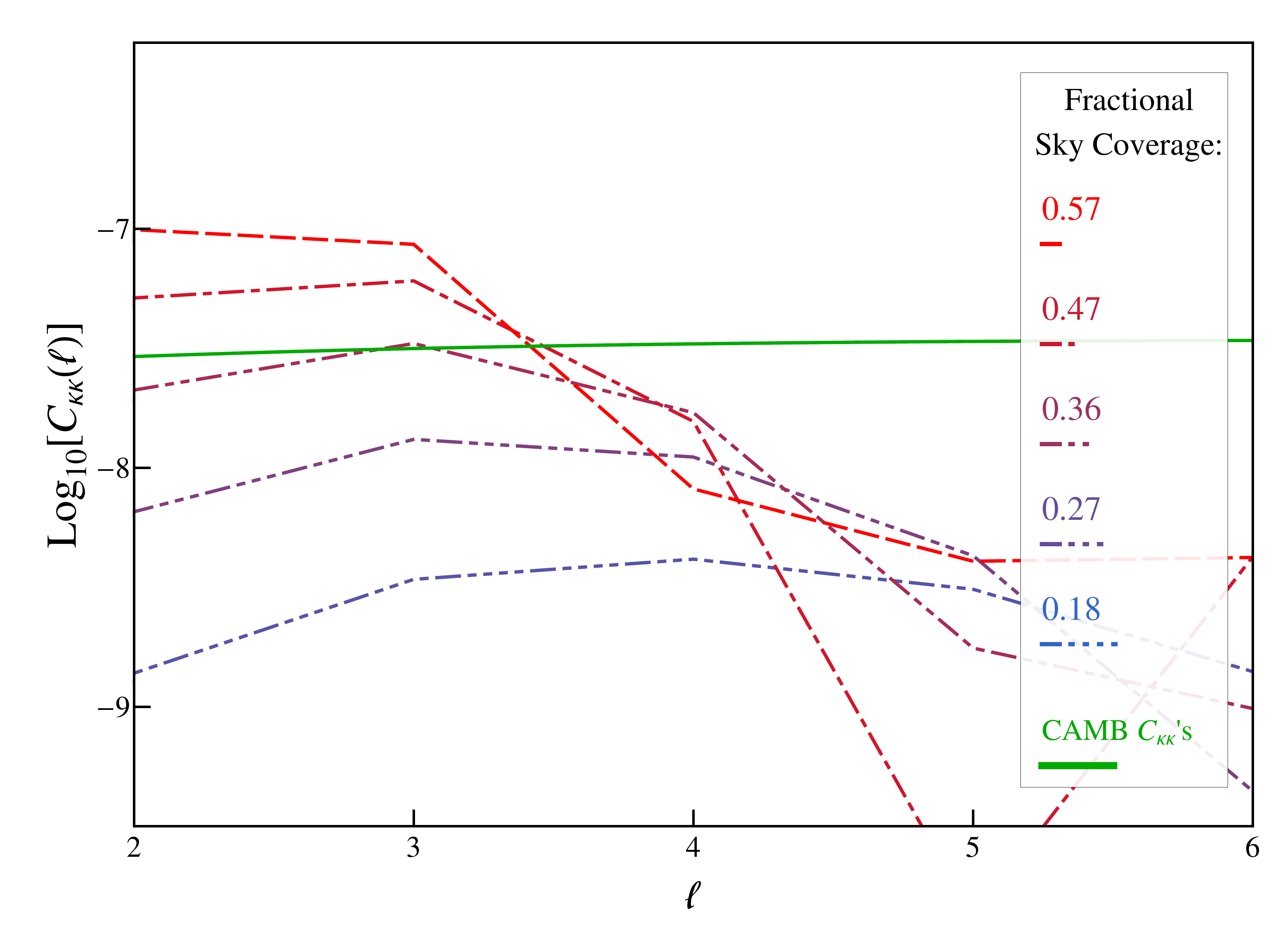}
	\caption{We plot the convergence power spectra purely due to boosting for various sky coverages
	for comparison to the full-sky primordial convergence power spectrum as generated by 
	{CAMB\sc}~\citep{Lewis:1999bs}. The side panel gives the fractional sky coverage for a survey
	for each of the plotted lines. Notice that when sky coverage becomes approximately a third of the sphere,
	the boosted power spectrum becomes larger than the expected primordial power
	spectrum. }\label{ps_plot}
\end{figure}

From above, the convergence due to a boost in the $-{\bf\hat{z}}$ direction 
(letting $\theta =\psi-\pi/2 $) is: 
\begin{equation}
\kappa\mbox{\ensuremath{\left(\theta,\phi\right)}}=\beta\cos\left(\theta\right)=2\beta\sqrt{\frac{\pi}{3}}Y_{1}^{0}(\theta,\phi)\,.
\end{equation}
Thus on a full-sky only a dipole contribution will be present in the
power spectrum:
\begin{eqnarray}
C^{\kappa\kappa}_{\ell} & =\frac{1}{(2\ell+1)}\sum_{m}|a_{lm}|^{2} & \Rightarrow\begin{cases}
C^{\kappa\kappa}_{1} &\!\!\!\!\! = \frac{4\pi}{9}\beta^{2}\\
C^{\kappa\kappa}_{\ell\neq1} & \!\!\!\!\! =0 .
\end{cases}\,\,\,
\end{eqnarray}
 On a cut sky however, there is  ``ringing,'' where power
from the dipole leaks into other moments, and instead 
\begin{equation}
a_{lm}=\intop_{\Omega}\kappa\left(\theta,\phi\right)Y_{l}^{m}(\theta,\phi)\mathrm{d}\Omega\,\,\, .
\end{equation}
The integral is taken over only the region of interest on the sky. In Figure~\ref{ps_plot}
we show the effect of increased sky coverage on the boosted convergence power spectrum for several specific
patches of sky, where the patches are chosen such that their center is located at $\frac{\pi}{2}$ from the 
boost axis. We see that experiments with larger sky coverage will be more sensitive to boosting effects,
and in fact for surveys which measure 36 per cent of the sky the boosted power spectrum may
be comparable to the expected primordial convergence power spectrum. We did find that
the power spectrum exhibited a dependence on the choice of sky coverage location in the 
$\theta$ direction, however calculations using approximately a third of the sky or above consistently became
larger than the expected full-sky primordial spectrum calculated from CAMB.

%%%%%%%%%%%%%%%%%%%%%%%%%%%%%%%%%%%%%
\section{Boost Effects on Reduced Shear}
\label{reduced}
%%%%%%%%%%%%%%%%%%%%%%%%%%%%%%%%%%%%%
Despite there being no change in the ellipticity of an image (as shown in sec.~\ref{peculiar}),
we want to investigate the effect of a boost on the reduced shear (eq.~\ref{rs}), since it includes a factor
of $\kappa$ which is altered by a boost.

If we consider an intrinsically circular background image with radius $R$ (which
we cannot measure directly) that is weakly lensed, we can measure the semi-minor and semi-major
axes of the image $a$ and $b$:
\begin{eqnarray}
a & = & \frac{R}{1-\kappa-\gamma},\\
b & = & \frac{R}{1-\kappa+\gamma}.
\end{eqnarray}
In the weak lensing limit, $a/b$ becomes
\begin{equation}
\frac{a}{b}=1+2\gamma .
\end{equation}

For the  magnification induced by boosting, 
the ratio $a/b$ will not change to first order in small quantities, because each
galaxy image is magnified symmetrically about its center. Thus
the magnitude of the shear, $\gamma$, will be affected by boosting only at second order. The relation between the shear due to  lensing alone ($\gamma_{WL}$) and the shear
due to weak lensing and boosting combined ($\gamma_{WL+\beta}$) is
\begin{equation}
\label{eqn:gammaWL}
\gamma_{WL}=\gamma_{WL+\beta}\left(1+\kappa_{\beta}\right).
\end{equation}
Here we have used the fact that to first order $\kappa_{WL+\beta}=\kappa_{WL}+\kappa_{\beta}$.
We can use  eqs.~\ref{rs} and~\ref{eqn:gammaWL} to infer that to first order in all
small parameters there is no change in the reduced shear
due to boost effects, so $g_{WL+\beta}=g_{WL}$.

%%%%%%%%%%%%%%%%%%%%%%%%%%%%%%%%%%%%%
\section{discussion}
\label{discussion}
%%%%%%%%%%%%%%%%%%%%%%%%%%%%%%%%%%%%%

With much attention being paid to weak lensing as a rich source of new
information about our universe, it is important to fully understand the 
challenges present for current and future experiments. With this work we have
characterized one particular systematic effect, the distortion of weak
lensing images due to the peculiar motion of our galaxy.

We have shown that, while ellipticities of galaxy images will not be exaggerated
due to boosting effects, the magnification of images will be changed at the 10 per cent
level. We have additionally shown that this effect can be neglected for high 
multipoles ($\ell\sim 5$ and above) as well as for surveys with small sky coverage.
However, as seen in Figure~\ref{ps_plot}, for surveys mapping a third of the 
sky, the convergence power spectrum purely from boost effects can become comparable to the expected 
primordial convergence power spectrum. This illustrates a need to account for the Lorentz
boost of weak-lensing images for large surveys such as LSST and DES. It could also affect measurements that probe the low-multipole weak-lensing signal, such as \citep{Kesden:2003zm}.   There are 
still several second order effects which should be investigated, in
particular the effect of boosting on isophotes and effects on flexion and other second
order lensing quantities.  

We noted that although circular images remain circular under a boost, isophotes 
(rings of constant intensity) would not necessarily follow the same behavior. 
The ability to neglect this effect will depend largely on differential 
magnification, relativistic aberration, and relativistic Doppler shift effects 
across an image.  The aberration and Doppler shifts themselves will depend 
on the radial luminosity profile and spectrum of a galaxy.

Higher order effects from boosting may become appreciable when considering 
second-order lensing effects, such as flexion.  To
higher order in small parameters, shear terms do 
appear: $\gamma_{1}=-\frac{\beta}{2}\delta\theta^{2}\sin\psi$.
Note that $\delta\theta$ is the angular size in radians of a galaxy, and 
will thus be at least an order of magnitude or two smaller than $\beta$.

It should also be noted that shear cannot be directly measured from observations.
Instead the reduced shear, $g$, is obtained, which is
insensitive to convergence effects to first order in small parameters.
In sec.~\ref{reduced} we showed that in this limit indeed measurements
of reduced shear would not be affected to first order and boost effects can 
safely be neglected.

\section*{Acknowledgments}
We would like to thank Tim Ivancic for his contributions to the early part of this work, Camille Bonvin for conversations regarding \citep{Bonvin:2008ni}, and the referee for pointing out the derivation at the beginning of section \ref{peculiar}.
AY and JM were supported in part by a US Department of Education GAANN grant to the CWRU Department of Physics.  AY was also supported by a NASA Earth and Space Science Fellowship -- Grant TRN507323.
GDS is supported in part by a grant from the US DOE to the particle astrophysics theory group at CWRU.

%%%%%%%%%%%%%%%%%%%%%%%%%%%%%%%%%%%%%
%\section{References}
%%%%%%%%%%%%%%%%%%%%%%%%%%%%%%%%%%%%%

%\bibliographystyle{h-physrev}
%\addcontentsline{toc}{section}{\refname}\bibliography{CMB_boost}

\end{document}